\documentclass[10pt,aps,prb,twocolumn,amsmath,amssymb,superscriptaddress]{revtex4-1}
\usepackage{amsmath}
\usepackage{graphicx}
\usepackage{color}
\usepackage{multirow}
\usepackage[caption = false]{subfig}

\def\beq{\begin{eqnarray}}
\def\eeq{\end{eqnarray}}
\def\beqq{\begin{eqnarray*} \color{blue} }
\def\eeqq{\end{eqnarray*}}

\def\V{\mathcal{V}}
\def\C{\mathcal{C}}

\def\Nd{{N_d}}
\def\Nddiff{{N_{d}^{\mathrm{diff}}}}

\begin{document}

\title{Semistochastic Heat-bath Configuration Interaction method: selected configuration interaction with semistochastic perturbation theory}
\author{Sandeep Sharma}
\email{sanshar@gmail.com}
\affiliation{Department of Chemistry and Biochemistry, University of Colorado Boulder, Boulder, CO 80302, USA}
\author{Adam A. Holmes}
\email{aah95@cornell.edu}
\affiliation{Department of Chemistry and Biochemistry, University of Colorado Boulder, Boulder, CO 80302, USA} 
\affiliation{Laboratory of Atomic and Solid State Physics, Cornell University, Ithaca, NY 14853, USA}
\author{Guillaume Jeanmairet}
\affiliation{Max Planck Institute for Solid State Research, Heisenbergstra{\ss}e 1, 70569 Stuttgart, Germany} 
\affiliation{Sorbonne Universités, UPMC Univ Paris 06, CNRS, Laboratoire PHENIX, F-75005 Paris, France}
\author{Ali Alavi}
\email{a.alavi@fkf.mpg.de}
\affiliation{Max Planck Institute for Solid State Research, Heisenbergstra{\ss}e 1, 70569 Stuttgart, Germany}
\affiliation{Dept of Chemistry, University of Cambridge, Lensfield Road, Cambridge CB2 1EW, United Kingdom}
\author{C. J. Umrigar}
\email{cyrusumrigar@gmail.com}
 \affiliation{Laboratory of Atomic and Solid State Physics, Cornell University, Ithaca, NY 14853, USA}
\begin{abstract}
We extend the recently proposed heat-bath configuration interaction (HCI) method [Holmes, Tubman, Umrigar,  \emph{J. Chem. Theory Comput.} {\bf 12}, 3674 (2016)], by introducing a
semistochastic algorithm for performing multireference Epstein-Nesbet perturbation theory, 
in order to completely eliminate the severe memory bottleneck of the original method.
The proposed algorithm has several attractive features.
First, there is no sign problem that plagues several quantum Monte Carlo methods.
Second, instead of using Metropolis-Hastings sampling, we use the Alias method to directly sample determinants
from the reference wavefunction, thus avoiding correlations between consecutive samples.
Third, in addition to removing the memory bottleneck, semistochastic HCI (SHCI) is faster than the deterministic variant for many
systems if a stochastic error of 0.1 mHa is acceptable.
Fourth, within the SHCI algorithm one can trade memory for a modest increase in computer time.
Fifth, the perturbative calculation is embarrassingly parallel.
The SHCI algorithm extends the range of applicability of the original algorithm, allowing us to calculate the
correlation energy of very large active spaces.
We demonstrate this by performing calculations on several first row dimers including F$_2$ with an active space of
(14e, 108o), Mn-Salen cluster with an active space of (28e, 22o), and Cr$_2$ dimer with up to a quadruple-zeta basis set
with an active space of (12e, 190o). For these systems we were able to obtain better than 1 mHa accuracy
with a wall time of merely 55 seconds, 37 seconds, and 56 minutes on 1, 1, and 4 nodes, respectively. 
\end{abstract}
\maketitle

\section{Introduction}

Many methods, e.g., coupled cluster and M{\o}ller-Plesset perturbation theory, can accurately and efficiently treat the electronic correlation of single-reference (weakly-correlated) systems.
In particular, coupled cluster with singles, doubles and perturbative triples (CCSD(T))
is very accurate for such systems and is often referred to as the ``gold standard" of quantum chemistry.
However, these methods fail catastrophically when applied to multireference (strongly-correlated) systems, such as molecules undergoing chemical
reactions or systems containing transition metal atoms with partially filled 
\emph{d} or \emph{f} orbitals.

One common approach for tackling such multireference problems is to abandon the Hartree-Fock wavefunction and
instead use a multideterminantal reference wavefunction obtained by correlating a subset of orbitals around the Fermi surface.
Examples include the complete active space (CAS) method, in which all possible occupancies of orbitals within the active space are included,
and the restricted/generalized active space (RAS/GAS) methods\cite{Olsen1988,Malmqvist1990,Ma2011a}, in which further restrictions are placed on
the occupancies of the active orbitals in order to reduce the size of the Hilbert space.
The CAS method is limited to about 16 active electrons and orbitals.
Other possibilities include highly accurate but approximate methods such as
the density matrix renormalization group (DMRG)~\cite{White1992,White1993,ChaSha-ARPC-11}, full configuration
interaction quantum Monte Carlo (FCIQMC)~\cite{BooThoAla-JCP-09,CleBooAla-JCP-10}, and its semistochastic improvement
(S-FCIQMC)~\cite{PetHolChaNigUmr-PRL-12},
which routinely treat up to about 40-50 active orbitals.
A well-chosen active space often results in a reference wavefunction that contains qualitatively correct
physics. However, quantitative accuracy requires one to take into account the dynamical correlation by allowing excitations
into inactive-space orbitals. Common methods for including dynamical correlation include multireference configuration
interaction (MRCI) and its size-consistent variants\cite{wernercaspt2,Knowles1992,Shamasundar2011}, various flavors of multireference perturbation theory\cite{caspt2,Fink2009,Angeli2001,Hirao1992}, and
multireference coupled cluster theories\cite{Lyakh2012,Evangelista2012,Sharma2016,Sharma2015mrlcc,jeanmairet17}.
The accuracy of these methods is often limited by the fact that only a relatively small number of active space orbitals can be used in the reference wavefunction because the cost of enlarging the active space increases exponentially with the number of orbitals.
 

Although the number of determinants in a CAS scales combinatorially with the number of active electrons and orbitals,
many of these determinants are ``configurational deadwood,'' and do not contribute appreciably to the reference wavefunction~\cite{IvaRue-TCA-01}.
The so-called \emph{selected configuration interaction} (SCI) methods~\cite{HurMalRan-JCP-73, BuePey-TCA-74, EvaDauMal-CP-83, Har-JCP-91, SteWenWilWil-CPL-94, WenSteWil-IJQC-96, IvaRue-TCA-01, Neese-JCP-03, AbrShe-CPL-05, BytRue-CP-09, Eva-JCP-14, Kno-MP-15, SchEva-JCP-16, TubLeeTakHeaWha-JCP-16, LiuHof-JCTC-16, CafAppGinSce-ARX-16},
which have been in use for more than four decades,
take advantage of this fact and generate a reference wavefunction by selecting only important determinants, rather than including all determinants in the CAS.
A subset of these methods improve upon the variational energy by employing a perturbative correction to the energy using multireference Epstein-Nesbet perturbation theory.
We refer to these methods as
\emph{selected configuration interaction plus perturbation theory} (SCI+PT) methods.
The first such
method was called \emph{configuration interaction perturbing a multi-configurational zeroth-order wavefunction selected iteratively} (CIPSI)~\cite{HurMalRan-JCP-73}.

The focus of this paper is a newly-introduced SCI+PT method called \emph{heat-bath configuration interaction} (HCI).
HCI~\cite{HolTubUmr-JCTC-16} distinguishes itself from other SCI+PT techniques by employing an algorithm that greatly improves the
efficiency of both the variational and perturbative steps.
Although it is more efficient than other SCI+PT methods, HCI, in its original formulation, is limited by
a memory bottleneck because it stores in memory
all the determinants that contribute to the perturbative correction (see the end of Section~\ref{HCI_alg} for more details). 

In this paper, we introduce a semistochastic implementation of multireference Epstein-Nesbet perturbation theory,
and use it to overcome the memory bottleneck of HCI. 
This method has several attractive properties.
First, it does not have a sign problem that plagues quantum Monte-Carlo methods.
Second, instead of using the Metropolis-Hastings method, we use the Alias method to sample the variational wavefunction directly, so the samples are all uncorrelated.
Third, in addition to removing the memory bottleneck, semistochastic HCI (SHCI) is often faster than the deterministic variant
if a stochastic error of 0.1 mHa is acceptable.
Fourth, within the SHCI algorithm one can trade memory for a modest increase in computer time.
Fifth, the perturbative calculation is embarrassingly parallel.

In Section~\ref{HCI_alg} we review the improvements made in the original HCI algorithm that make it much more
efficient than other SCI+PT algorithms.
In Section~\ref{stoch_PT}, we present our stochastic perturbation theory which removes the memory bottleneck
of the original HCI algorithm, and then our semistochastic perturbation theory which is more efficient
in terms of computer time than either the deterministic or the stochastic variants.
In Section~\ref{implementation} we provide various implementation details of both the variational and the
perturbative parts of our algorithm.
We then demonstrate the utility of the stochastic and semistochastic methods by applying them in Section~\ref{results} to various diatomic molecules
including F$_2$ with an active space of (14e, 108o), Mn-Salen cluster with an active space of (28e, 22o),
and Cr$_2$ dimer with up to a quadruple-zeta basis set with an active space of (12e, 190o), obtaining
energies that are accurate to better than 1 mHa 
with very modest computer resources.
Finally, in Section~\ref{conclusion}, we conclude and discuss future research directions.

\section{Heat-bath Configuration Interaction}
\label{HCI_alg}
We begin by describing the HCI algorithm in its original formulation~\cite{HolTubUmr-JCTC-16}, emphasizing
the key innovations that make it much more efficient than other SCI+PT methods. In the following discussion the indices $i,j,\cdots$ will be used for determinants in the variational space $\V$ and the indices $a,b,\cdots$
will be used for determinants in $\C$, the space of determinants that are connected to $\V$ but not in $\V$.
Similar to other SCI+PT methods, HCI has two stages: (1) a variational stage, in which a variational wavefunction
is obtained as a linear combination of a set of determinants chosen by an iterative procedure, and (2) a perturbative
stage, in which the second-order correction to the variational energy is computed using
multireference Epstein-Nesbet perturbation theory~\cite{EPS-PR-26,Nes-PRS-55},
but each stage is much faster than in other SCI+PT methods.

\subsection{Variational Stage}
At the start of the algorithm, $\V$ consists
of some initial set of determinants, usually just the Hartree-Fock determinant.
Then, at each iteration, new determinants are added to $\V$, chosen using a parameter $\epsilon_1$,
as follows.
The initial wavefunction is the ground state of the Hamiltonian in $\V$,
$|\Psi_0\rangle=\sum_i c_i |D_i\rangle$.
At each iteration:
\begin{enumerate}
\item Add to the variational space $\V$, all determinants $D_a$ in the space of connections $\C$, such that
\begin{align}
\left | H_{ai} c_i\right| > \epsilon_1 \label{eq:eps1}
\end{align}
for at least one determinant $D_i$ in the current $\V$.
\item Calculate the lowest eigenvalue $E_0$ with eigenvector $|\Psi_0\rangle=\sum_i c_i |D_i\rangle$ of the
Hamiltonian in $\mathcal{V}$.
\end{enumerate}

The iterations are terminated when the number of new determinants is less than a threshold, e.g., 1\% of the current size of $\V$,
or when a maximum number of iterations is reached.
Since the values of $c_i$ tend to be larger in the initial iterations when there are few determinants in $\V$, $\epsilon_1$ is
set during the first few iterations to be larger than its final value.

HCI takes advantage of the fact that the double excitation matrix elements depend only on the four orbitals
whose occupancy is changing.
Step 1 is performed efficiently by storing the double excitation matrix elements in order of decreasing magnitude,
so that no time is wasted on determinants that do not meet the cutoff in Eq.~\ref{eq:eps1}. For details, we
refer the reader to the original HCI paper~\cite{HolTubUmr-JCTC-16}. Thus, we see that HCI
identifies new determinants to add to $\V$ in a manner that is more efficient than other SCI methods
in two ways:
\begin{itemize}
\item  HCI uses a selection criterion which is \emph{cheap to evaluate} for each determinant, namely Eq.~\ref{eq:eps1}.
In contrast, other SCI methods use a criterion based on a perturbative expression which is more expensive; for
example, CIPSI~\cite{HurMalRan-JCP-73} uses the magnitude of the coefficient of the first-order correction to the wavefunction, namely
$\left|\frac{\sum_i H_{ai} c_i}{E_0-H_{aa}}\right|>\epsilon_1$.
\item  HCI evaluates its selection criterion (Eq.~\ref{eq:eps1}) \emph{only for those doubly-excited determinants which will be
added to $\V$!}
By comparison, other SCI methods iterate through \emph{all} candidate determinants $\{D_a\}$ (determinants
for which there exists at least one nonzero matrix element $H_{ai}$ with $D_i\in \V$), evaluating their
expensive selection criteria for each one.
\end{itemize}

The simplification in HCI is possible because it was demonstrated~\cite{HolTubUmr-JCTC-16} that variation in the perturbative expression
for the coefficients is dominated by variation in the largest-magnitude term in the numerator, since the matrix
elements $\{H_{ai}\}$ and coefficients $\{c_i\}$ span many orders of magnitude.
The minor deviation from optimality in the choice of the most important determinants is by far outweighed by the fact that
many more determinants can be included because the selection criterion of HCI allows the variational and perturbative steps
to be performed at much reduced computational cost.

\subsection{Perturbative Stage}
The variational wavefunction is used to define the zeroth order Hamiltonian, $H_0$ and the perturbation, $V$,
\begin{align}
H_0 &= \sum_{i,j} H_{ij} |D_i\rangle\langle D_j| + \sum_{a } H_{aa} |D_a\rangle\langle D_a|. \nonumber\\
V &= H - H_0\label{eq:part}
\end{align}
It can easily be verified that $|\Psi_0\rangle$ is the ground state of $H_0$ with eigenvalue $E_0$.
Using the partitioning in Eq.~\ref{eq:part}, the first-order correction of the wavefunction $|\Psi_1\rangle$ and the second-order
energy correction $\Delta E_{2}$ can be written as
\begin{align}
 |\Psi_1\rangle &= \frac{1}{E_0- H_0}V|\Psi_0\rangle\nonumber\\
 & = \sum_{a} \frac{\sum_{i}H_{ai} c_i}{E_0 - E_a}  |D_a\rangle
 \end{align}
and
 \begin{align}
 \Delta E_{2} & = \langle\Psi_0|V|\Psi_1\rangle\nonumber\\
  &=\sum_{a} \frac{\left(\sum_{i}H_{ai} c_i\right)^2}{E_0 - E_a}\label{eq:PTb} ,
 \end{align}
where $E_a=H_{aa}$.
It is worth noting that the expression for the total energy, $E_0 + \Delta E_{2}$ is identical to that for the mixed estimator of the energy
used in quantum Monte Carlo calculations, provided that the projected wavefunction is replaced by the perturbed wavefunction.

This expression in Eq.~\ref{eq:PTb} is expensive to calculate, as it requires a summation over many small terms.
Instead, HCI includes only those terms in the sum that contribute substantially,
 \begin{align}
 \Delta E_{2}\approx\sum_{a} \frac{\left(\sum_{i}^{\left(\epsilon_2\right)}H_{ai} c_i\right)^2}{E_0 - E_a},
\label{eq:PTc}
 \end{align}
where $\sum^{\left(\epsilon_2\right)}$ denotes a sum in which all terms in the sum that are smaller in magnitude than
$\epsilon_2$ are discarded, i.e., $\sum_i^{\left(\epsilon_2\right)} H_{ai}c_i$  includes only terms for which
$\left\vert H_{ai} c_i \right\vert > \epsilon_2$.

Once again, since the double excitation matrix elements are stored in order of decreasing magnitude,
no time is spent on the doubly-excited terms that do not contribute to the sum.
The parameter $\epsilon_2$ is kept much smaller than the parameter $\epsilon_1$ because discarding small amplitude determinants can lead to significant errors in the calculation of dynamical correlation.
In the original HCI paper~\cite{HolTubUmr-JCTC-16}, for each $\epsilon_1$, several values of $\epsilon_2$ were used, and the energy for
$\epsilon_2=0$ was obtained by extrapolation.  In this paper, a single value of $\epsilon_2$ is used, that is sufficiently
small to recover the $\epsilon_2=0$ limit to a precision that is better than 1 mHa.

It was shown in the previous publication~\cite{HolTubUmr-JCTC-16} that the above algorithm is highly efficient and can be
used to obtain sub-milliHartree accuracy for challenging problems like all-electron (48e, 42o) Cr$_2$ with
the small Ahlrichs double-zeta basis~\cite{schafer1992fully}
in a few minutes on a single computer core.
However, since the contributions from all $i$ in Eq.~\ref{eq:PTc} must be summed and then squared, the efficient deterministic
approach to computing the perturbative correction requires storing the partial sums $\left\{\sum_i^{\left(\epsilon_2\right)}H_{ai}c_i\right\}$ for all $a$ for which
$\left\vert H_{ai} c_i \right\vert > \epsilon_2$ which creates a severe memory bottleneck. 

\begin{figure}
\begin{center}
\includegraphics[width=0.5\textwidth]{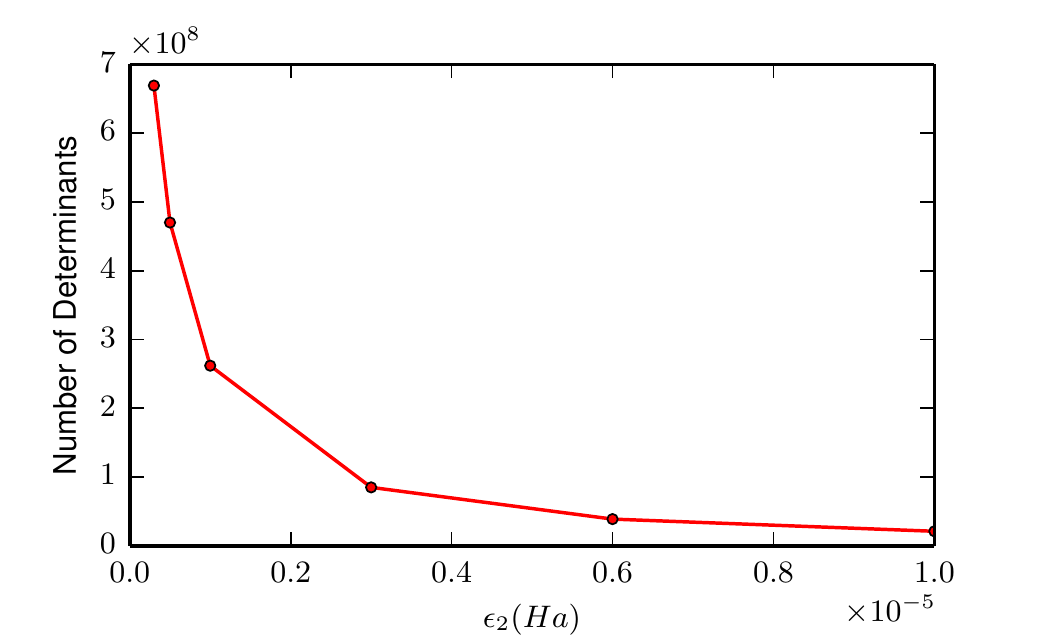}
\end{center}
\caption{Demonstration that the number of determinants in the perturbative space $\C$ increases rapidly as $\epsilon_2$ is reduced. These calculations, for the C$_2$ dimer with a QZ basis, used $\epsilon_1=2\times 10^{-4}$ Ha in the variational calculation resulting in 403071 variational determinants. All the calculations presented in the results section use an $\epsilon_2=10^{-8}$ Ha, whereas the smallest $\epsilon_2$ used in this graph is $3\times 10^{-7}$ Ha.
}\label{fig:numpt}
\end{figure}

To see this, we note that the $N_v$ determinants in the variational space are connected to ${\cal
  O}(n^2v^2N_v)$ determinants in the perturbative space with nonzero
  Hamiltonian matrix elements, where $n$ is the number of electrons and $v$ is
  the number of virtual orbitals. For a relatively conservative number of
  $n=12$, $v=50$ and $N_v=10^7$, the perturbative space will contain over
  $10^{12}$ determinants, requiring over 10 terabyte memory. The original HCI
  algorithm reduces this storage requirement by orders of magnitude by
  storing only determinants $D_a$ for which
  $\left|H_{ai}c_i\right|>\epsilon_2$ for at least one determinant $D_i \in
  \mathcal{V}$. But it should be recognized that using a large $\epsilon_2$ to truncate the perturbative space can lead to significant errors in the perturbative correction. If on the other hand $\epsilon_2$ is reduced to decrease the error in the perturbative correction the number of determinants in the perturbative space $\C$ rapidly increases as shown in Figure~\ref{fig:numpt}, making the calculation infeasible.

In the following section we show that this memory bottleneck can be completely eliminated, without having to increase $\epsilon_2$, by using a stochastic or semistochastic version of the perturbation theory.

\section{Stochastic Multireference Perturbation Theory}
\label{stoch_PT}

We first discuss the stochastic method for computing the perturbative correction before discussing the more efficient semistochastic method.

\subsection{Stochastic PT}

We write the perturbative correction in a slightly different form than presented in Eq.~\ref{eq:PTc} to highlight the fact that it is a bilinear function of the coefficients of the zeroth-order state.
\begin{align}
\Delta E_{2} &= \sum_{a} \frac{1}{E_0 - E_a}\left(\sum_{ij}^{\left(\epsilon_2\right)}H_{ai}H_{aj} c_ic_j\right)
\label{eq:stoch}.
\end{align}

We compute the expected value of this expression stochastically by employing $N_s$ samples,
each of which consists of $\Nd$ determinants $\left\{D_i\right\}$ sampled from $\mathcal{V}$ with probability
\beq
p_i = \frac{|c_i|}{\sum_i |c_i|}.
\eeq
Any given sample will contain $\Nddiff$ distinct determinants $D_i$ with some number of repeats $w_i$, such that \[\sum_{i}^{\Nddiff}w_i = \Nd\]
The number of repetitions, $w_i$, is distributed according to the well-known multinomial distribution.  The mean and second moment, for $i \ne j$, of this distribution are
\begin{align}
\label{expec_wi}
\langle w_i\rangle =& p_i\Nd\\
\label{expec_wij}
\langle w_i w_j \rangle =& p_ip_j\Nd(\Nd-1),
\end{align}
where $\left\langle\cdot\right\rangle$ denotes the expectation value of a quantity evaluated for a sample of
$N_d$ determinants, a notation we will use hereafter.

Using these expressions, the unbiased estimate of the second-order perturbation can be calculated from the sampled wavefunction as follows,
%
\begin{widetext}
\begin{align}
\Delta E_{2}
=& \sum_{a} \frac{1}{E_0 - E_a}\left[\sum_{ij}^{\V} H_{ai}H_{aj} c_ic_j\right]\nonumber\\
=& \sum_{a} \frac{1}{E_0 - E_a}\left[\sum_{i\neq j}^{\V} H_{ai}H_{aj} c_ic_j + \sum_{i}^{\V} H_{ai}^2 c_i^2\right]\nonumber\\
=& \left\langle \sum_{a} \frac{1}{E_0 - E_a}\left[\sum_{i\neq j}^{\Nddiff} \frac{w_i w_j c_ic_j H_{ai}H_{aj}}{\langle w_i w_j \rangle} + \sum_{i}^{\Nddiff} \frac{w_i c_i^2 H_{ai}^2}{\langle w_i \rangle}\right] \right\rangle \nonumber\\
=& \left\langle \sum_{a} \frac{1}{E_0 - E_a}\left[\sum_{i\neq j}^{\Nddiff} \frac{w_i w_j c_ic_j H_{ai}H_{aj}}{p_ip_j\Nd(\Nd-1)} + \sum_{i}^{\Nddiff} \frac{w_i c_i^2 H_{ai}^2}{p_i\Nd}\right] \right\rangle \nonumber\\
=& \frac{1}{\Nd(\Nd-1)} \left\langle \sum_{a} \frac{1}{E_0 - E_a} \left[\left(\sum_{i}^{\Nddiff} \frac{ w_i  c_i H_{ai}}{p_i}\right)^2  +\sum_{i}^{\Nddiff} \left(\frac{w_i(\Nd-1)}{p_i } - \frac{ w_i^2}{p_i^2}\right)c_i^2 H_{ai}^2\right] \right\rangle,
\label{eq:stofinal}
\end{align}
\end{widetext}
where for brevity we have suppressed the superscript $\left(\epsilon_2\right)$ on the $i$ and $j$ sums,
though of course we will always use a nonzero $\epsilon_2$ value for efficiency. 

Going from the $2^{nd}$ to the $3^{rd}$ line above, we replace the sum over the states in $\V$ by a sum
over the sample, so in order to have an unbiased expectation value we divide the two terms
by $\langle w_i w_j \rangle$ and $\langle w_i \rangle$ respectively.
In going from the $3^{rd}$ to the $4^{th}$ line we use Eqs.~\ref{expec_wi} and \ref{expec_wij}.

In practice, the exact average in Eq.~\ref{eq:stofinal} will be replaced by an average over $N_s$ samples.
For any $\Nd \ge 2$ we obtain an unbiased estimate of the second-order correction to the energy and this estimate can be made progressively more precise by averaging over a large number of samples $N_s$.
Each batch contains an independently chosen set of $\Nddiff$ determinants and thus there is no autocorrelation between consecutive batches.
This is in sharp contrast to discrete-space quantum Monte Carlo methods, such as the FCIQMC method~\cite{BooThoAla-JCP-09,CleBooAla-JCP-10}
and its semistochastic improvement~\cite{PetHolChaNigUmr-PRL-12}, for which the autocorrelation time increases
both with system size and the size of the basis to the point that it can become difficult to accurately
estimate the statistical error.  This drawback of the FCIQMC method is ameliorated but not eliminated by using the
more efficient sampling method of Ref.~\onlinecite{HolChaUmr-JCTC-16}.

We note that the expression in Eq.~\ref{eq:stofinal} is evaluated in much the same way as the
deterministic evaluation of the perturbative correction using a single batch, the main difference being that the $N_v$ variational
determinants have been replaced by the much smaller subset of $\Nddiff$ distinct sampled determinants and
that an additional summation is needed to ensure that the result is unbiased.
Note that for each sample,
the summation over $a$ in Eq.~\ref{eq:stofinal} is restricted to only those determinants in $\mathcal{C}$ that have a nonzero Hamiltonian matrix element with the $\Nddiff$ determinants used to sample the zeroth-order wavefunction. 

\begin{figure}[h]
\subfloat[]
{\includegraphics[width=0.5\textwidth]{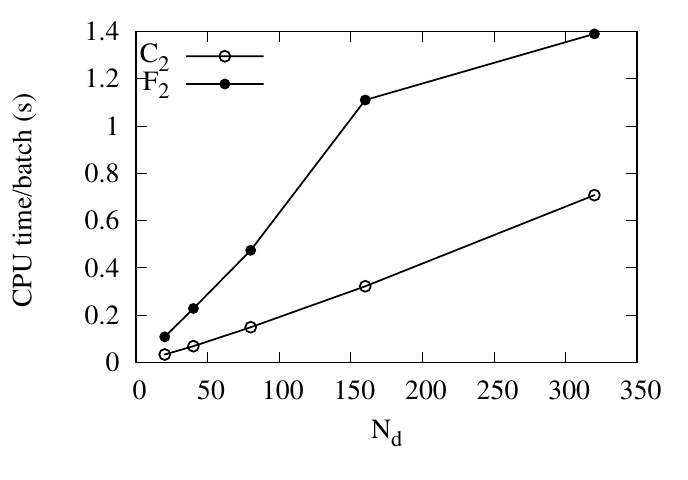}\label{fig:cpuscaling}}\\
\subfloat[]
{\includegraphics[width=0.5\textwidth]{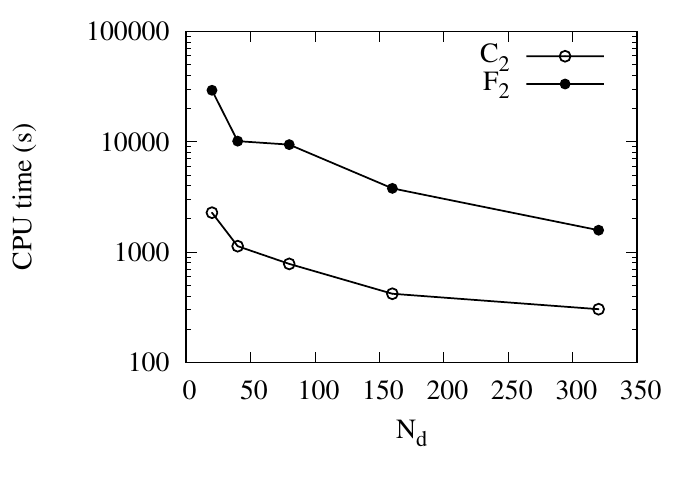}\label{fig:cpuscaling2}}
\caption{a) Demonstration of the near-linear scaling of the CPU time per batch for the perturbative calculation as a function of the number of determinants, $\Nd$, sampled in each batch.
The open and the filled circles are for the C$_2$ and F$_2$ molecules respectively with cc-pVQZ basis sets.
b) CPU time in seconds required to reach a standard deviation of 0.1 mHa for various values of $\Nd$.
Note the initial rapid decrease in CPU time followed by a more gradual decrease at larger $\Nd$ values.
}
\end{figure}

Figure~\ref{fig:cpuscaling} shows that the CPU time per sample increases nearly linearly with $\Nd$, the number of determinants in the sample, for the C$_2$ and F$_2$ molecules. As shown in Section~\ref{implementation}, the scaling contains two terms: one that scales linearly with $\Nd$ and another that scales as $\Nd \log(N_d)$ (footnote.~\onlinecite{log_corr}).
Figure~\ref{fig:cpuscaling2} shows the CPU time necessary to reach a standard deviation of less than 0.1 mHa versus the number of determinants in the sampled wavefunction $\Nd$.
There is a rapid initial decrease followed by a much shallower decrease beyond about $\Nd = 200$.
Consequently, it is desirable to use as large a value of $\Nd$ as memory allows.
Another consideration is that $N_s$ needs to be large enough to get a reasonable estimate of the statistical error,
and since the computer time is approximately $\propto \Nd N_s$, it sometimes makes sense to use a smaller $\Nd$
than available memory allows.
In all the calculations presented in Section~\ref{results}, we have used $\Nd = 200$, even though the computer time
could be greatly reduced by using a larger $\Nd$ for the larger systems.

It is worth mentioning that the memory bottleneck can also be removed without recourse to the stochastic method.
This can be achieved by dividing the $N_v$ determinants in $\V$ into $N_b$ batches, each containing
on average $N_d$ determinants ($N_b = N_v / N_d)$, 
and computing the contribution from all pairs of batches independently.
In Section~\ref{implementation} we will see that the leading cost of performing the calculation for each pair of batches is $\propto  N_d$ and so the cost of performing the entire calculation containing $N_b$ batches scales as
$N_d N_b^2 \propto N_v^2/N_d$.
Thus, for $N_v > N_d$, the cost of the deterministic approach scales quadratically with $N_v$.
In contrast, the cost of the stochastic approach scales sub-linearly with $N_v$ because additional low weight determinants are sampled only infrequently.
The price to be paid is that the computed energy has a stochastic error.
Still, for large $N_v$ and small $\epsilon_2$, the $N_s$ required to get a statistical error of 1 mHa in the stochastic method is much smaller than the
$N_b^2$ required in the deterministic calculation, making the stochastic approach the more efficient choice.


\subsection{Semistochastic PT}

Although the stochastic method eliminates the memory bottleneck, for large variational spaces and large basis sets it takes many stochastic samples, $N_s$, to reduce the stochastic error to the desired value.
The computer time can be reduced by using a semistochastic method in which the perturbative calculation is split into two steps.
In the first step a deterministic perturbative calculation is performed using a relatively loose threshold $\epsilon_{2}^{\mathrm{\rm d}}$ to obtain the deterministic second-order correction $\Delta E_2^D[\epsilon_{2}^{\mathrm{\rm d}}]$. The error incurred due to the use of large $\epsilon_{2}^{\mathrm{\rm d}}$ in the deterministic calculation can be corrected stochastically by calculating the difference between the second-order energies obtained from a tight threshold $\epsilon_{2}$ and a loose threshold $\epsilon_{2}^{\mathrm{\rm d}}$:
\begin{align}
\Delta E_2[\epsilon_{2}] = (\Delta E_2^S[\epsilon_{2}] - \Delta E_2^S[\epsilon_{2}^{\mathrm{d}}]) + \Delta E_2^D[\epsilon_{2}^{\mathrm{d}}], \label{eq:semisto}
\end{align} 
where $\Delta E_2^S[\epsilon_{2}]$ and $\Delta E_2^S[\epsilon_{2}^{\mathrm{d}}]$ are the second-order energies calculated with $\epsilon_{2}$ and $\epsilon_{2}^{\mathrm{d}}$ respectively using the stochastic method.
The key point here is that both the  $\Delta E_2^S[\epsilon_{2}]$ and $\Delta E_2^S[\epsilon_{2}^{\mathrm{d}}]$ are calculated using the same set of sampled $N_d$ variational determinants and thus there is substantial cancellation of stochastic error, and almost no increase in memory or computer time.
The value of $\epsilon_{2}^{\mathrm{d}}$ affects the statistical error of the energy for given computer time, but not the expectation value of the energy.
In the results section we show that using the semistochastic method for Cr$_2$ dimer can speed up the calculation by more than a factor of 2.
This speed up can be even larger if the available computer memory permits a smaller $\epsilon_{2}^{\mathrm{d}}$ to be used.
In the limit that one can afford to use $\epsilon_{2}^{\mathrm{d}} = \epsilon_{2}$, the stochastic noise is completely eliminated. Thus the semistochastic method gives us the ability to go smoothly from the fully deterministic to the fully stochastic algorithm.

\section{Implementation}
\label{implementation}
Here we briefly describe the implementation and the leading-order cost of the various steps of the algorithm.
In the variational stage there are three main operations, identifying the significant determinants to be included in the variational space, building the Hamiltonian matrix and diagonalizing the matrix.
The cost of identifying important determinants is $O(kN_v \ln(N_v) + kN_v \ln(N_p))$, where $N_v$ is the number of determinants
in the variational space,
$k$ is the average number of $H_{ai}$ elements for determinants $D_i$ in $\V$ that satisfy Eq.~\ref{eq:eps1},
and $N_p=k N_v$ is the total number of new determinants that satisfy the criterion in Eq.~\ref{eq:eps1}.
The two terms in the cost function result from generating $kN_v$ determinants and then doing a binary search
of the list of the $N_v$ existing variational determinants and the $N_p$ newly generated determinants before
including the determinant just generated in the newly generated determinant list.

In the current implementation we store all the nonzero elements of the Hamiltonian in memory using a \emph{list of lists} (LIL) sparse storage format.
In LIL format for each row we store a list containing the column index and the value of the nonzero Hamiltonian matrix elements.

The determinant labels are bit-packed strings that represent the occupancies of the up-spin ($\alpha$) and
the down-spin ($\beta$) orbitals.
To build the Hamiltonian efficiently, we first generate a list of all unique $\beta$ strings and associated with each $\beta$ string we store a list of all determinants in $\V$ that have that $\beta$ string.
We also generate a list of all unique $\alpha$ strings with $N_{\alpha}-1$ electrons and associated with each $\alpha$ string we store a list of all determinants in $\V$ that give the $\alpha$ string on removing one $\alpha$ electron.
Here, $N_\alpha$ is the number of $\alpha$ electrons in our system.
Determinants that are related to each other by double or single $\alpha$ excitations have the same $\beta$ string, and all the pairs of determinants that are related to each other with the remaining possible single or double excitations have the same $\alpha$ string with $N_\alpha -1$ electrons.
Hence to find the connected determinants, only the determinants in these two lists need to be considered
rather than the entire set of $N_v$ determinants in $\V$.
Once the Hamiltonian is generated the Davidson algorithm is used to diagonalize it and the most expensive step there is the Hamiltonian wavefunction multiplication which costs $O(k N_v)$.
Despite the fact that the Hamiltonian is sparse, building it is the most expensive part of the variational step, and storing it is currently the largest memory bottleneck in the code.
In the future we intend to implement the \emph{direct} method~\cite{Knowles1984,IvaRue-TCA-01} for carrying out Hamiltonian wavefunction
multiplication which does not require storing the Hamiltonian and can take less computer time
as well. 


The stochastic perturbation step has two major components: sampling $\Nd$ determinants from the list of $N_v$ variational determinants, and identifying the determinants in $\C$ that are connected to these $\Nd$ determinants
and computing their contribution to the perturbative correction, Eq.~\ref{eq:stofinal}.
The $\Nd$ determinants are sampled  using the Alias method~\cite{walker1977efficient,kronmal1979alias}, which has an initial one-time memory cost of $O(N_v)$, and a subsequent cost of $O(\Nd)$ each time a sample is drawn.
The Alias method was used by some of us~\cite{HolChaUmr-JCTC-16} for efficiently sampling determinants in the S-FCIQMC method.
The computer time for identifying the connected determinants along with their contributions is $O(n^2v^2\Nd \log(n^2v^2\Nd) + n^2v^2\Nd \log(N_v))$, while the memory required is $O(n^2v^2\Nd + N_{v})$.
Since the minimum required value of $\Nd$ is just two, the memory requirement for the stochastic perturbation
theory is smaller than that of other parts of the calculation.

We have parallelized the entire code using hybrid OMP/MPI (open multiprocessing/message passing interface) programming to make full use of the symmetric multiprocessor (SMP) architecture of most modern computers. A separate MPI process is initiated on each computer node and then each process forks into several threads (one for each computational core) on the node.
The variational wavefunction is replicated on each node but a single copy is shared among the different threads on a node.
As mentioned previously, the most memory intensive data structure is the LIL used to store the sparse Hamiltonian matrix.
The rows of the Hamiltonian are distributed in round-robin fashion among the nodes.  Of course, only the nonzero elements are stored.
With this strategy the storage of the Hamiltonian and the computation of the Hamiltonian wavefunction multiplication is distributed approximately evenly between the different nodes and threads.
The perturbative step is embarrassingly parallel and no special strategy is needed to parallelize this step.

\section{Benchmarks}
\label{results}
We perform frozen core calculations (including all the Hartree-Fock virtual orbitals in the active space) on a series of first row dimers including C$_2$, N$_2$, O$_2$, NO and F$_2$ with cc-pVDZ, cc-pVTZ and cc-pVQZ basis sets.
(In the interest of brevity, these three basis sets will sometimes be abbreviated as DZ, TZ, and QZ, respectively.)
Although the active spaces used for the first row diatomics have large Hilbert spaces, they do not exhibit strong correlation in their equilibrium geometry and traditional methods such as CCSD(T) are cheap and reliable.
Thus, we also perform frozen-core calculations on the Cr$_2$ dimer using cc-pVDZ, cc-pVTZ and cc-pVQZ bases, which have active spaces containing (12e, 68o), (12e, 118o) and (12e, 190o) respectively.
Cr$_2$ is well known for being very strongly correlated.  Most multi-reference methods can use no more than
the minimal active space and therefore fail to get even qualitatively correct dissociation curves.
Finally, we also perform calculations on the Mn-Salen model complex which is a prototypical strongly correlated inorganic molecule containing open shell \emph{d}-orbitals giving rise to nearly degenerate singlet and triplet ground states.
For all the systems we obtain energies that are accurate to 1 mHa for the chosen basis;
for the first-row dimers we compare to S-FCIQMC energies~\cite{Cleland2012}, for the Cr$_2$ dimer we perform internal
convergence tests since there are no good approximations to the FCI energies in the literature, and for Mn-Salen we compare to DMRG energies~\cite{Sharma2016b}.

\subsection{First row diatomics}
\begin{table*}[htb]
\caption{Ground state SHCI energies of the C$_2$, N$_2$, O$_2$, NO and F$_2$ molecules with bond lengths of 1.2425, 1.0977, 1.2075, 1.1508 and 1.4119 \AA\ respectively and DZ, TZ and QZ basis sets.
The variational cutoff, $\epsilon_1$, the number of determinants in the variational space, $N_v$, the variational energy, Var, and the total energy, Total, are shown.
The statistical error in the last digit is shown in parentheses.
All perturbative calculations were performed with an $\epsilon_2 = 10^{-8}$ Ha and $\Nd = 200$. The last three rows show that the SHCI calculations converge with much looser $\epsilon_1$ threshold when MP2 natural orbitals, rather than canonical HF orbitals, are used.
The final three columns show the wall time in seconds required to perform the variational and the stochastic perturbative parts of the calculation. All the results obtained from perturbation theory (PT) agree within error bars with the results published previously using FCIQMC~\cite{Cleland2012}.
Each calculation was performed on a single node (see text for details).}\label{tab:diatomics}
\begin{tabular}{lccrrcccrrr}
\hline
\hline
& & & & &\multicolumn{2}{c}{Energy (Ha)}&& \multicolumn{3}{c}{Wall time (sec)}\\
\cline{6-7}
\cline{9-11}
Molecule&~~Basis~~&~~Sym~~&~~~~$\epsilon_1$ (Ha)~~~~&~~$N_{v}$~~&~~~~Var~~~~&~~~Total~~~&&Var&PT&Total\\
\hline
C$_2$& DZ & $^1$A$_{1g}$ & $5\times 10^{-4}$&28566&     -75.7217&       -75.7286(2)&&   1&      2&3 \\
C$_2$& TZ & $^1$A$_{1g}$ & $3\times 10^{-4}$&142467     &-75.7738       & -75.7846(3)&& 7&     4&11 \\
C$_2$& QZ & $^1$A$_{1g}$ & $2\times 10^{-4}$&403071&    -75.7894&       -75.8018(4)&&   36&      10&46\\
\\
N$_2$& DZ & $^1$A$_{1g}$ & $5\times 10^{-4}$&   37593   &-109.2692&     -109.2769(1)&&  1&2&3 \\
N$_2$& TZ & $^1$A$_{1g}$ & $3\times 10^{-4}$&   189080& -109.3608&      -109.3748(6)&&  10&4&14 \\
N$_2$& QZ & $^1$A$_{1g}$ & $2\times 10^{-4}$&499644&    -109.3884       &-109.4055(9)   &&44&10&54\\
\\
O$_2$& DZ & $^3$A$_{1g}$ & $5\times 10^{-4}$&   52907&  -149.9793&      -149.9878(2)&&  2&3&5\\
O$_2$& TZ & $^3$A$_{1g}$ & $3\times 10^{-4}$&   290980& -150.1130&      -150.1307(8)&&  17&6&23 \\
O$_2$& QZ & $^3$A$_{1g}$ & $2\times 10^{-4}$&   770069& -150.1541&      -150.1748(9)&&  72&23&95\\
\\
NO& DZ & $^2$B$_{1}$ & $5\times 10^{-4}$&       48305&  -129.5881       &-129.5997(3)&& 2&3&5 \\
NO& TZ & $^2$B$_{1}$ & $3\times 10^{-4}$&227004&        -129.6973&      -129.7181(9)&&  18&7&25\\
NO& QZ & $^2$B$_{1}$ & $2\times 10^{-4}$&       606381& -129.7311&      -129.7548(9)&&  93&23&116\\
\\
F$_2$& DZ & $^1$A$_{1g}$ & $5\times 10^{-4}$&   68994&  -199.0913&      -199.1001(7)&&  2&3&5\\
F$_2$& TZ & $^1$A$_{1g}$ & $3\times 10^{-4}$&   395744& -199.2782&      -199.2984(9)&&  8&6&14\\
F$_2$& QZ & $^1$A$_{1g}$ & $2\times 10^{-4}$&1053491&   -199.3463       &-199.3590(9)&& 129&22&151\\
\\
\multicolumn{8}{l}{\emph{Natural Orbitals}}\\
F$_2$& DZ & $^1$A$_{1g}$ & $1\times 10^{-3}$&16824&     -199.0871&      -199.0994(4)&&  0&3&3\\
F$_2$& TZ & $^1$A$_{1g}$ & $5\times 10^{-4}$&141433&    -199.2787&      -199.2972(7)&&  7&6&13\\
F$_2$& QZ & $^1$A$_{1g}$ & $5\times 10^{-4}$&221160&    -199.3355&      -199.3590(9)&&  28&27&55\\
\hline
\hline
\end{tabular}
\end{table*}

In the variational calculations we start with a value of $\epsilon_1$ during the first few iterations that
is larger than its final value because the values of $c_i$ tend to be larger in the initial iterations when there are few determinants in $\V$.
For example, for the cc-pVQZ basis set, we successively reduce the value of $\epsilon_1$ from $10^{-3}$ to $5\times 10^{-4}$ to $3\times 10^{-4}$ and $2\times 10^{-4}$ Ha, and perform 3 iterations at each value.
The cost of performing the first iteration at a value of $\epsilon_1$ is larger than that for subsequent iterations because relatively few new determinants are introduced after the first iteration.

Table~\ref{tab:diatomics} shows benchmark calculations on the first row dimers using a single node containing two Intel\textsuperscript{\textregistered} Xeon\textsuperscript{\textregistered} E5-2680 v2 processors of 2.80 GHz each and 128 gigabyte memory.  Among these calculations F$_2$ had the largest active space containing 14 electrons in 108 orbitals (14e, 108o) with a Hilbert space containing over $10^{20}$ determinants.  On a single node it required less than 3 minutes to get the energy converged to better than 1 mHa. It is not possible to perform the calculations for the larger systems and basis sets on a single node with the original algorithm because the cost of storing all the determinants in the space of
connections $\C$ that contribute to the perturbative corrections is prohibitive.
Interestingly, not only is the memory requirement of the stochastic method smaller than that of the original deterministic algorithm, but even the computer time required
to obtain sub-mHa accuracy is smaller, for the systems where the deterministic algorithm is feasible.

As expected, these calculations can be done even more efficiently, if the Hartree-Fock orbitals are replaced by natural orbitals from some approximate correlated theory.
For example the last three rows of Table~\ref{tab:diatomics} show that the calculations on the F$_2$ dimer, for all three basis sets, can be run with a larger $\epsilon_1$ resulting in about a factor of 3 speedup when MP2 natural orbitals are used.

\subsection{Cr$_2$ dimer}

\begin{table*}[htbp]
\caption{Ground state SHCI energies of the Cr$_2$ molecule with bond length 1.68 \AA\ with DZ, TZ and QZ basis sets.
For each basis set, calculations were performed with both the stochastic and the semistochastic methods described in Section~\ref{stoch_PT}.
The various columns in the table have the same meaning as those in Table~\ref{tab:diatomics}, but there are three new columns.
The column labeled $\epsilon_{2}^{\rm d}$ gives the value of the $\epsilon_2$ used to perform the deterministic calculation in the semistochastic variant of the method.
The column labeled PT$_{\rm det}$ shows the time required for the determinstic part of the PT calculation for the semistochastic method, and the column labelled PT(1) shows the time it would take to perform the stochastic perturbative step if we terminated the calculation after obtaining an uncertainty of 1 mHa.
The final column labeled \#Nodes shows the number of computer nodes used.
The extrapolated energies are obtained from a quadratic fit to the energies for $\epsilon_1 \le 20\times 10^{-5}$ Ha.
The relatively slow convergence of the TZ basis is due to an integrals file that uses poor quality natural orbitals from FCIQMC.
The energies in the TZ' rows were obtained using the same basis, but with
approximate natural orbitals obtained from an SHCI variational wavefunction.
They converge much faster.
The wall times are not shown for some calculations because these were done on a different computer architecture so a meaningful comparison is not possible.
}\label{tab:cr2}
\begin{tabular}{ccrcrcllrrrrc}
\hline
\hline
& & & & &\multicolumn{2}{c}{Energy (Ha)}&& \multicolumn{4}{c}{Wall time (sec)}&\\
\cline{6-7}
\cline{9-12}
~~Basis~~&~~Sym~~&~~~~$\epsilon_1$ (Ha)~~~~&~~~~$\epsilon_{2}^{\rm d}$ (Ha)~~~~&~~~~~~$N_{v}$~~~~~~&~~~~~~Var~~~~~~&~~~~~~Total~~~~~~~&&Var&PT$_{\rm det}$&PT(1)&\#Nodes\\ 
\hline
DZ & $^1$A$_{1g}$ & $8\times 10^{-5}$& $-$              & 3114163&-2099.4692&  -2099.4875(5)&&  395&      0&	983&1\\[2mm]
DZ & $^1$A$_{1g}$ &$50\times 10^{-5}$& $5\times 10^{-6}$&  210421&-2099.4344&  -2099.4851(1)&&  $-$&    $-$&	$-$&$-$\\
DZ & $^1$A$_{1g}$ &$20\times 10^{-5}$& $5\times 10^{-6}$&  832196&-2099.4560&  -2099.4864(1)&&  $-$&    $-$&	$-$&$-$\\
DZ & $^1$A$_{1g}$ & $8\times 10^{-5}$& $5\times 10^{-6}$& 3114163&-2099.4692&  -2099.48754(3)&&	395&    100&    246&1\\ 
DZ & $^1$A$_{1g}$ & $6\times 10^{-5}$& $5\times 10^{-6}$& 4708713&-2099.4724&  -2099.48782(7)&&	$-$&    $-$&    $-$&1\\ 
DZ & $^1$A$_{1g}$ & $5\times 10^{-5}$& $5\times 10^{-6}$& 6114463&-2099.4741&  -2099.48788(7)&&	$-$&    $-$&    $-$&1\\ 
DZ & $^1$A$_{1g}$ & $4\times 10^{-5}$& $5\times 10^{-6}$& 8390964&-2099.4760&  -2099.48809(3)&&  $-$&    $-$&    $-$&1\\ 
DZ & $^1$A$_{1g}$ & Extrapolated     & $-$              & $-$    & $-$      &  -2099.4887(2) &&	$-$&    $-$&   	$-$&$-$ \\
\\
TZ & $^1$A$_{1g}$ & $8\times 10^{-5}$& $-$              & 6268840&-2099.5051&  -2099.5276(7)&&  593&      0&	783&4\\
TZ & $^1$A$_{1g}$ & $7\times 10^{-5}$& $-$              & 7651680&-2099.5070&  -2099.5283(6)&&	862&      0&	828&4\\
TZ & $^1$A$_{1g}$ & $6\times 10^{-5}$& $-$              & 9666032&-2099.5090&  -2099.5283(6)&& 1024&      0&	795&4\\[2mm]
TZ & $^1$A$_{1g}$ &$50\times 10^{-5}$& $5\times 10^{-6}$&  379428&-2099.4637&  -2099.52183(1)&&	$-$&	$-$&	$-$&$-$\\
TZ & $^1$A$_{1g}$ &$20\times 10^{-5}$& $5\times 10^{-6}$& 1561960&-2099.4887&  -2099.52532(5)&&	$-$&	$-$&	$-$&$-$\\
TZ & $^1$A$_{1g}$ &$10\times 10^{-5}$& $5\times 10^{-6}$& 4496674&-2099.5017&  -2099.5277(1)&&	$-$&    $-$&   	$-$&$-$\\ 
TZ & $^1$A$_{1g}$ & $8\times 10^{-5}$& $5\times 10^{-6}$& 6268840&-2099.5051&  -2099.5283(1)&&	593&    250&   	233&4\\ 
TZ & $^1$A$_{1g}$ & $7\times 10^{-5}$& $5\times 10^{-6}$& 7651680&-2099.5070&  -2099.52851(4)&&	862&    319&   	313&4\\ 
TZ & $^1$A$_{1g}$ & $6\times 10^{-5}$& $5\times 10^{-6}$& 9687009&-2099.5090&  -2099.52884(4)&&1024&    394&   	328&4\\ 
TZ & $^1$A$_{1g}$ & $5\times 10^{-5}$& $5\times 10^{-6}$&12759006&-2099.5113&  -2099.52936(7)&&	$-$&    $-$&   	$-$&$-$ \\
TZ & $^1$A$_{1g}$ & Extrapolated     & $-$              & $-$    & $-$      &  -2099.5312(15) &&	$-$&    $-$&   	$-$&$-$ \\
\\
TZ'& $^1$A$_{1g}$ &$50\times 10^{-5}$& $5\times 10^{-6}$&  365104&-2099.4650&  -2099.52831(1)&&	$-$&	$-$&	$-$&$-$\\
TZ'& $^1$A$_{1g}$ &$20\times 10^{-5}$& $5\times 10^{-6}$& 1549370&-2099.4899&  -2099.52961(2)&&	$-$&	$-$&	$-$&$-$\\
TZ'& $^1$A$_{1g}$ &$10\times 10^{-5}$& $5\times 10^{-6}$& 4429824&-2099.5028&  -2099.53073(4)&&	$-$&    $-$&   	$-$&$-$\\
TZ'& $^1$A$_{1g}$ & $8\times 10^{-5}$& $5\times 10^{-6}$& 6185301&-2099.5062&  -2099.53104(3)&&	$-$&    $-$&   	$-$&$-$\\
TZ'& $^1$A$_{1g}$ & $7\times 10^{-5}$& $5\times 10^{-6}$& 7553766&-2099.5081&  -2099.53124(5)&& $-$	$-$&    $-$&$-$\\
TZ'& $^1$A$_{1g}$ & $6\times 10^{-5}$& $5\times 10^{-6}$& 9510287&-2099.5101&  -2099.53135(6)&&$-$&    $-$&   	$-$&$-$\\
TZ'& $^1$A$_{1g}$ & $5\times 10^{-5}$& $5\times 10^{-6}$&12479803&-2099.5123&  -2099.53155(4)&&	$-$&    $-$&   	$-$&$-$\\
TZ'& $^1$A$_{1g}$ & Extrapolated     & $-$              & $-$    & $-$      &  -2099.5325(4) &&	$-$&    $-$&   	$-$&$-$\\
\\
QZ & $^1$A$_{1g}$ & $8\times 10^{-5}$& $-$              & 9516339&-2099.5246&  -2099.5553(13)&&	1584&	   0&	3578&4\\[2mm]
QZ & $^1$A$_{1g}$ &$50\times 10^{-5}$&$5\times 10^{-6}$ &  497747&-2099.4782&  -2099.5562(3)&&	 $-$&	 $-$&	 $-$&$-$\\
QZ & $^1$A$_{1g}$ &$20\times 10^{-5}$&$5\times 10^{-6}$ & 2285120&-2099.5064&  -2099.5560(3)&&	 $-$&	 $-$&	 $-$&$-$\\
QZ & $^1$A$_{1g}$ &$10\times 10^{-5}$&$5\times 10^{-6}$ & 6768521&-2099.5208&  -2099.55649(3)&&	 $-$&	 $-$&	 $-$&$-$\\
QZ & $^1$A$_{1g}$ & $8\times 10^{-5}$&$5\times 10^{-6}$ & 9516339&-2099.5246&  -2099.55670(7)&&	1584&	 372& 	1429&4\\ 
QZ & $^1$A$_{1g}$ & $6\times 10^{-5}$&$5\times 10^{-6}$ &14812200&-2099.5290&  -2099.55668(8)&&	 $-$&    $-$& 	 $-$&$-$\\
QZ & $^1$A$_{1g}$ & $5\times 10^{-5}$&$5\times 10^{-6}$ &19481471&-2099.5315&  -2099.55682(8)&&	4960&   2751& 	1087&4\\
QZ & $^1$A$_{1g}$ & Extrapolated     & $-$              & $-$    & $-$      &  -2099.5571(3) &&	$-$&    $-$&   	$-$&$-$ \\
\hline
\hline
\end{tabular}
\end{table*}

The Cr$_2$ dimer is well known to be a very challenging system for most electronic structure methods.
We perform frozen core calculations by including all the virtual orbitals in the active space with a bond length of 1.68 \AA\ using the cc-pVDZ-DK, cc-pvTZ-DK and cc-pVQZ-DK basis sets.
The relativistic effects are included using the second-order Douglas-Kroll-Hess Hamiltonian.
For all the basis sets calculations are performed using natural orbitals obtained by first performing a short unconverged FCIQMC calculation.
For the TZ basis, calculations are also performed using
approximate natural orbitals obtained from an SHCI variational wavefunction.
The active spaces with the DZ, TZ and QZ basis sets contained (12e, 68o), (12e, 118o) and (12e, 190o) respectively, with the largest Hilbert space containing more than $10^{21}$ determinants.

\begin{figure}[htbp]
\begin{center}
\includegraphics[width=0.48\textwidth]{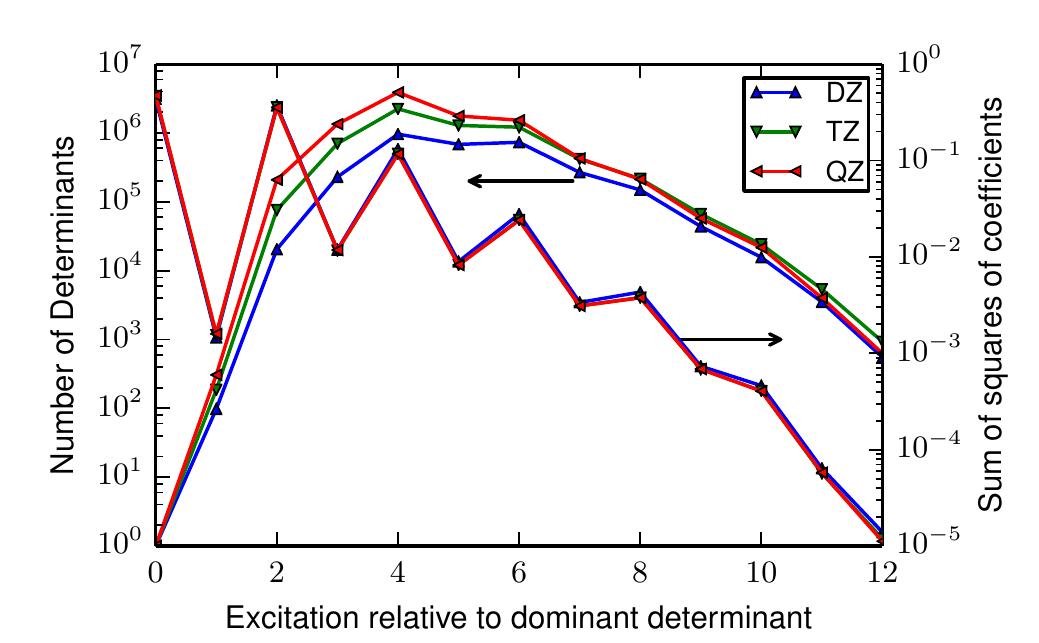}
\end{center}
\caption{The number of determinants and the sum of the squares of the coefficients of the
variational wavefunction versus excitation level relative to the dominant determinant
for each of the three basis sets, with $\epsilon_{1}=8\times 10^{-5}$ Ha.
Determinants of all excitation orders are important.}
\label{fig:excitation_order}
\end{figure}

Figure~\ref{fig:excitation_order} shows the number of determinants and the sum of the squares of the coefficients of the
variational wavefunction for each excitation level relative to the dominant determinant 
for the three basis sets with $\epsilon_{1}=8\times 10^{-5}$ Ha.
The wavefunction contains determinants with excitation orders all the way up to the maximum possible of 12.
Hence CI expansions that are truncated at double or even quadruple excitations are far from adequate.

Table~\ref{tab:cr2} shows the variational and the total energies for each of the three basis sets.
Although the variational energies are far from convergence, the total energies including the perturbative corrections converge rapidly as $\epsilon_1$ is reduced.
Also shown in the table are the computer times for those calculations that employed the computer architecture described above.
The cost of the perturbative calculation for a constant statistical error is smaller for the semistochastic variant
compared to the stochastic variant of the method.
This is particularly important when a small statistical error is required since the stochastic error of the perturbative calculation decreases as $1/\sqrt{N_s}$,
where $N_s$ is the number of samples used in the stochastic perturbative calculation.
For example, for the TZ basis with $\epsilon_1=7\times 10^{-5}$ Ha, if a statistical error of 0.1 mHa is required the computer time is
82800 sec for stochastic PT but only 394+32800=33194 sec for semistochastic PT.
The efficiency of the semistochastic method can be further improved by reducing the value of $\epsilon_{2}^{\mathrm{d}}$.
For large systems it is most efficient to use the smallest value of $\epsilon_{2}^{\mathrm{d}}$ that is permitted by computer memory, though
for small systems a larger value of $\epsilon_{2}^{\mathrm{d}}$ can be optimal.

The computed total energies for the smallest $\epsilon_1$ values for the DZ and QZ bases is within 1 mHa of their respective extrapolated energies
but not for the TZ basis.  Hence, as described above, we also calculated energies using approximate natural orbitals obtained from an SHCI wavefunction.
These calculations are denoted by TZ' (although the basis is the same) and they are much better converged.

It is noteworthy that the cost of the perturbative correction relative to the variational calculations decreases as the size of the variational space $\V$ increases.
In fact, the CPU time required to reach a fixed statistical error is rather insensitive to the size of the variational space for a given basis set.
This is because as the variational space is increased, the perturbative correction becomes smaller, and because the additional variational determinants have relatively few connections that satisfy the $\epsilon_2$ threshold.
This indicates that much larger calculations could be performed if the memory and computer time of the variational step were reduced, potentially by using the direct CI method\cite{Knowles1984}.

\subsection{Mn-Salen}
\begin{figure}
\begin{center}
\includegraphics[width=0.4\textwidth]{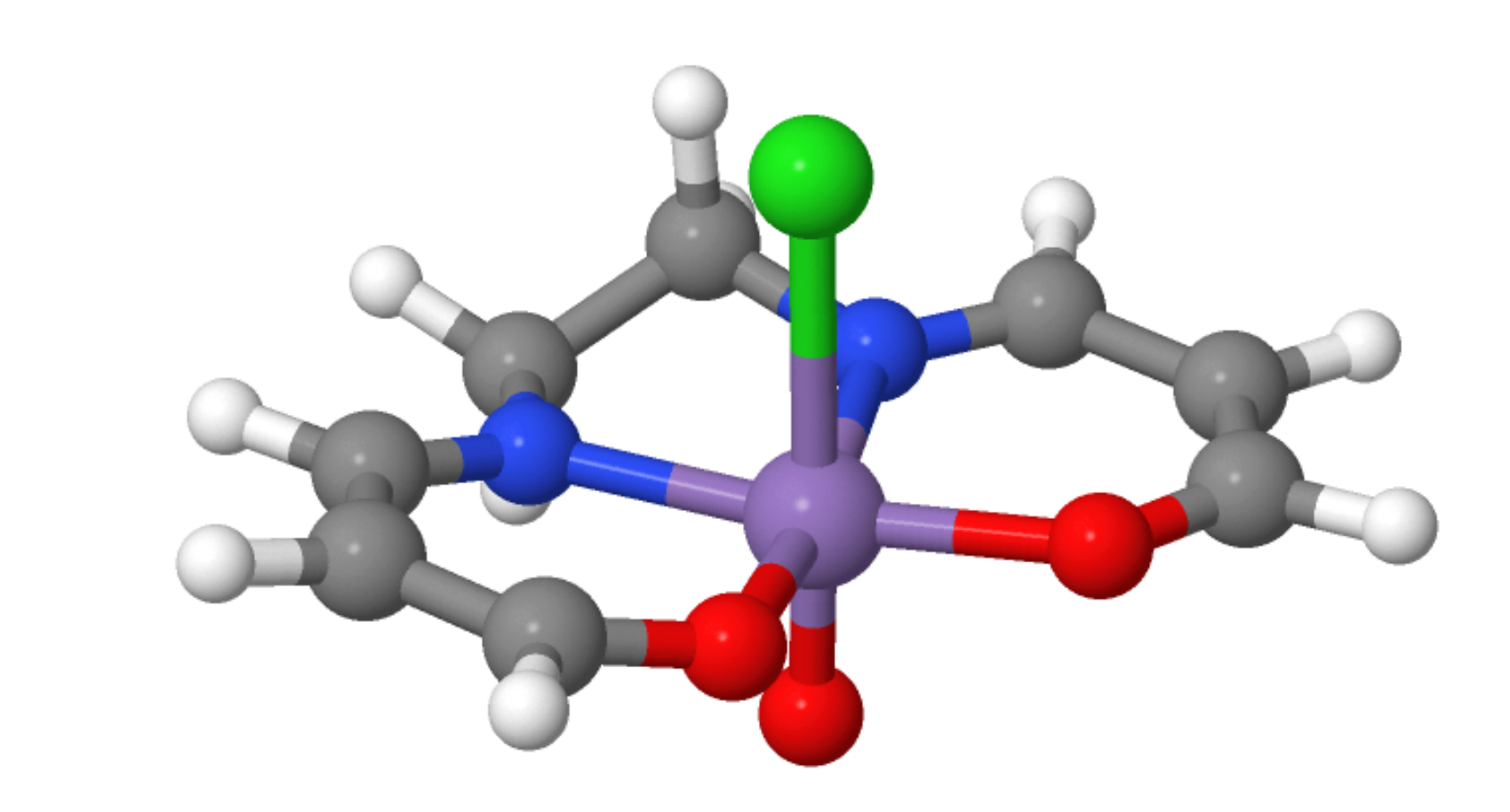}
\end{center}
\caption{The geometry of the chlorine containing neutral oxo-Mn(salen) cluster was optimized with (10e, 10o)-CASSCF and a 6-31G* basis by Ivanic et al.~\cite{Ivanic2004}. The Mn, Cl, N, O, C, H atoms are shown in purple, green, blue, red, grey and white respectively.}\label{fig:Mnsalen}
\end{figure}

The calculations for the first-row dimers and the Cr$_2$ dimer were for the exact frozen core energies.
We now
demonstrate that SHCI can be used to calculate the active space energy of a prototypical strongly correlated molecule like Mn-Salen (MnClO$_3$N$_2$C$_8$H$_{10}$) (see Figure~\ref{fig:Mnsalen}) very quickly.
Mn-Salen derivatives such as Jacobson's catalyst are used to catalyze enantioselective epoxidation of olefins.
Despite their widespread use and importance, the mechanism of the catalysis reaction is not known and has spawned a series of theoretical studies\cite{C4CP00721B,Stein2016,Bogaerts2015,Ma2011,Sears2006,Ivanic2004,Wouters2014,Linde1999,Abashkin2001,Abashkin2004,Abashkin2001a}.
Recently, some of us performed DMRG-SCF calculations~\cite{Sharma2016b} on the model cluster with the cc-pVDZ basis set using an active space of (28e, 22o).
The initial orbitals were obtained by using the HOMO-13 to LUMO+7 canonical Hartree Fock orbitals, which were subsequently optimized using the DMRG-SCF method. Here we perform the SHCI calculations on the converged orbitals obtained at the end of the converged DMRG-SCF calculations.
The results in Table~\ref{tab:mnsalen} show that both the singlet and the triplet energies converge to better than 1 mHa accuracy in only 37 seconds on a single node.

\begin{table}
\caption{Comparison of the DMRG and SHCI energies (E+2251) Ha of the singlet and triplet states of Mn-Salen. DMRG was performed with an $M=2000$ and SHCI was performed with $\epsilon_1=2\times 10^{-4}$ Ha, $\epsilon_2=1\times 10^{-8}$ Ha and $\Nd=200$. The wall time needed to perform the SHCI calculation on a single computer node is shown in the final column(see text for additional details).}\label{tab:mnsalen}
\begin{tabular}{cccccc}
\hline
\hline
& &\multicolumn{3}{c}{Energy (Ha)}&\\
\cline{3-5}
Sym&$N_v$&Var&PT&DMRG&Wall time (s)\\
   &     &   &  &Ref.~\onlinecite{Sharma2016b}& \\
\hline
$^1$A&232484&-0.7880&-0.7980(7)&-0.7991&37\\
$^3$A&208334&-0.7910&-0.7994(8)&-0.8001&32\\
\hline
\end{tabular}
\end{table}

\section{Conclusions}
\label{conclusion}

We have introduced a stochastic and semistochastic implementations of multireference Epstein-Nesbet perturbation theory,
for computing the expectation value of the perturbative correction to the variational energy
of a multi-determinant wavefunction without storing all the contributing determinants.
In addition to completely removing the memory bottleneck, the semistochastic algorithm is faster than the fully deterministic algorithm
for most systems if a stochastic noise of 0.1 mHa is acceptable.

Our method is capable of efficiently computing
the correlation energies of very large active spaces, as we have demonstrated by computing the energies
of the challenging, multireference systems Mn-Salen (28e, 22o) and Cr$_2$ (12e, 190o).
For all systems studied we obtained correlation energies accurate to within 1 mHa.
In the case of the first-row dimers and Mn-Salen we compared to FCIQMC and DMRG energies in the literature.
For Cr$_2$ there are no published values, but one of the positive features of our method
is that one can reliably check the convergence within the method itself.

Having removed the memory bottleneck in the perturbative step, the largest memory requirement comes from storing
the Hamiltonian in the variational space. The next step is to create an efficient method
for obtaining the variational wavefunction without storing the Hamiltonian.
Other research directions include the optimization of the orbitals within the CAS space, and the calculation of excited states.

\begin{acknowledgements}
The calculations made use of the facilities of the Max Planck Society's Rechenzentrum Garching. SS acknowledges the startup package from the University of Colorado.  AAH and CJU were supported in part by NSF grant ACI-1534965. CJU thanks Garnet Chan for valuable discussions.
\end{acknowledgements}


\begin{thebibliography}{10}
\bibitem{log_corr}
The logarithmic corrections come from having to do binary searches to check
  whether a determinant is already present in the variational space, or, in the
  space of connected determinants that have already been generated.
\end{thebibliography}
\providecommand{\refin}[1]{\\ \textbf{Referenced in:} #1}

\end{document}